# Focus Point on Tensions in Cosmology from Early to Late Universe: Part II: New Directions in the Light of Observations from the Most Modern Astronomical Facilities

S.Capozziello[1,2,3], E. Di Valentino[4], V.G.Gurzadyan[5]

[1] Dipartimento di Fisica E. Pancini, Universit di Napoli Federico II", via Cinthia 9, I-80126, Napoli, Italy,
[2] INFN, Sezione di Napoli, via Cinthia 9, I-80126 Napoli, Italy,
[3] Scuola Superiore Meridionale, Largo S. Marcellino 10, I-80138, Napoli, Italy,
[4] School of Mathematical and Physical Sciences, University of Sheffield, Hounsfield Road, Sheffield S3 7RH, United Kingdom
[5] Center for Cosmology and Astrophysics, Alikhanian National Laboratory and Yerevan State University, 0025, Yerevan, Armenia.

**Abstract.** The papers included in this Focus Point collection are devoted to the studies on the cosmological tensions and challenges stimulated by the latest observational data. The first results of the LARES-2 laser ranging satellite on the high precision testing of the frame-dragging effect predicted by General Relativity are presented. The data on the S-stars monitoring in the Galactic center obtained by GRAVITY collaboration were analysed within the Physics-informed neural network (PINN) approach. The results enabled to probe the role of the cosmological constant, of the dark matter, the star cluster in the core of the Galaxy obtaining an upper limit for the star density. The topics include the conversion of high-frequency relic gravitational waves into photons in cosmological magnetic field, cosmological gravitational waves stochastic background generation through the spontaneous breaking of a global baryon number symmetry, observational predictions of the Starobinsky inflation model and other studies.

The Hubble tension and other cosmological tensions continue to challenge the theoretical studies and stimulate further the ongoing observational surveys [1–6]. Theoretical studies include variety of modifications of gravity, dark matter and dark energy models, up to attempts to revise certain basic physical principles and the search of genuine difference between the late and early Universe.

The papers included in this Focus Point collection are devoted to the studies on the cosmological tensions stimulated by the latest observational data obtained by James Webb Space Telescope, GRAVITY collaboration monitoring the S-stars in the Galactic center, NANOGrav collaboration, on LARES 2 laser ranging satellite on high precision testing of General Relativity.

The first results of the LARES 2 space experiment to test General Relativity regarding the frame-dragging effect are presented [7]. Laser ranging satellite LARES 2 (LAres RElativity Satellite 2) was successfully launched in July 2022 from Kourou spaceport (French Guiana), with the new launch vehicle VEGA C of Italian Space Agency (ASI), European Space Agency and space propulsion company AVIO. The results are shown to be in complete agreement with the predictions with General Relativity. It is stated that, the conceptual simplicity of the LARES 2 experiment as compared with those conducted by means of the LARES and LAGEOS satellites provides a significant advance in high precision tests of General Relativity. The paper is dedicated to John Archibald Wheeler, a major figure of fundamental physics of the XXth century. In [8] the results of computation of 110 Earth's tidal perturbation significant modes in Doodson number classification for the parameters of LARES 2 satellite are presented. The data obtained by LARES and LARES 2 satellites are used to constrain the modified gravity models proposed to deal with the nature of the dark sector and cosmological tensions.

NANOGrav collaboration results on the detection of stochastic long wavelength gravitational waves have increased the interest to the anticipated properties of the primordial gravitational waves which might survive since the inflationary epoch. The role of the conversion of high-frequency relic gravitational waves into photons at presence of a cosmological magnetic field is studied in [9]. For gravitational waves in the sub-horizon regime during radiation dominance epoch it is shown the significant increase of relic gravitational waves with the frequencies $k > 10^{-11}$ Hz at magnetic field strength $B \approx 1$ nGs. While tensor gravitational waves propagating in the magnetic field generate photons, which in turn generate scalar metric perturbations, electromagnetic and scalar fields with corresponding components of the energy-momentum tensor compensate each other. As a result, the amplitude of the tensor gravitational waves remains unchanged. The interaction of electromagnetic waves with the primary plasma during radiation dominated epoch is also studied.

The Starobinsky inflation model is considered in [10], aiming to derive solutions involving the observables, the scalar spectral index $n_s$, the tensor-to-scalar ratio $r$, along with the consequences for reheating, e-fold number. The equation linking inflation with reheating is used to derive the spectral index $n_s$. Agreement of the Starobinsky model with the measurements of the power spectrum of primordial curvature perturbations and the bounds on the spectrum of primordial gravitational waves is concluded. The reheating is also considered in [11] for two classes of inflationary models, generalized $\alpha$-attractor and the $\alpha$-Starobinsky models. It is shown that the maximum value of the reheating temperature limits the overproduction of gravitinos during reheating. A universal scaling behavior for the reheating temperature is stated in both those inflationary models, outlining the importance of reheating parameters in the link with the observables.

The properties of the primordial gravitational waves which might survive since the inflationary epoch are studied in [12], and it is shown that gravitational waves during the radiation-dominated phase of the evolution of the Universe the gravitational waves can get their properties smeared. Namely, the primordial gravitational waves due to hyperbolicity at their propagation through the matter inhomogeneities can get flattened their primordial spectrum, with possible links to CMB properties, e.g. [13]. This effect can influence the features of the primordial gravitational waves as of the goal of ongoing experimental surveys.

Generation of early cosmological stochastic gravitational waves through the spontaneous breaking of a global baryon number symmetry for a Majorana neutron is considered in [14]. A complex scalar field associated with a baryonic Majoron or Baryo-Majoron undergoing a strong first-order phase transition (FOPT) induced by thermal corrections from a new heavy fermion field. The considered model also introduces a new pseudo Nambu-Goldstone boson as a possible candidate for dark matter and predicts baryon first-order phase transitions detectable in upcoming gravitational wave NANOGrav-type experiments.

S-star dynamics in the vicinity of the supermassive black hole in the center of the Milky Way is known to act as a laboratory for testing General Relativity, probing modified gravity models. The Physics-informed neural network (PINN) approach was used to analyse the S-star data to trace the role of the dark matter, i.e. of an extended configuration. and of the dynamics of S-stars in the vicinity of the supermassive black hole in the Galactic center is performed within General Relativity treatment. It is shown that the PINN training does not detect any perturbation on the dynamics of S2 star by an extended gravitational structure up to 0.01% of the total mass inside the apocenter of S2 [15]. Such constraints are useful at the analysis of the dark halos of galaxies (e.g. [16]) and the dark matter models. Then, the S2 star observational data, again via PINN, were used to probe the star cluster in the core of the Galaxy, studying the scattering of S2 star on stars of the cluster. Described by a random force given by the Holtsmark distribution acting on S2 star, an upper limit was obtained for the star density of the core cluster, $n_{crit} \approx 8.3 \times 10^6 pc^{-3}$ [17]. Note, the PINN, first used in [15,17] to probe modified gravity, the dark matter configurations and the stellar core in the Galactic center, is then further applied in [18].

5D Brans–Dicke theory is used in [19] to describe the baryogenesis and primordial light element formation and to explain the accelerated expansion on the Universe without involving matter fields in 5D or dark energy in 4D. The connection with dark matter relic abundance is discussed. In [20] the Vlasov–Einstein equations are used to describe the dynamics and the evolution of gravitating structures, the accelerated expansion of the universe with the cosmological constant and the cosmological tensions. Traversable wormhole solutions are investigated in [21] determined by an exponential shape function and fractional redshift function for four f(R) models including the Starobinsky model. The wormhole solutions are found which do not require exotic matter.

The ongoing observations with modern facilities are expected to provide further precious information on the cosmological tensions and hence will be decisive in their clarification and the choice of proper theories and models.

**Data availability** This paper has no associated data or the data will not be deposited. There is no data because all obtained results are in the paper. All authors confirm that there is no part of the paper that requires data.